# How to transform, with a capacitor, thermal energy into usable work.

**E. N. Miranda**[1]
CONICET – CCT Mendoza
5500 – Mendoza, Argentina
and
Facultad de Ingeniería
Universidad de Mendoza
5500 – Mendoza, Argentina

**Abstract:**

The temperature dependence of the dielectric permittivity is taken into account to study the energy change in a capacitor that follows a cycle between a cold and a hot thermal reservoirs. There is a net energy gain in the process that, in principle, can be transformed into usable work. The article is simple enough as to be used with keen undergraduates that have taken a university general physics or thermodynamics course. Further experimental work and a possible technological application are suggested.

---

[1] E-Mail: emiranda@mendoza-conicet.gov.ar

# 1. Introduction

In dealing with capacitors and dielectrics the [usual] texts [1, 2, 3] used for a general physics course consider the dielectric constant to be independent of temperature. Afterwards, if the students take a course in electromagnetism [4, 5, 6] they learn a microscopic description of dielectrics that take into account the polarizability of molecules, and the well-know equations of Clausius-Mossotti and Langevin- Debye are obtained. The latter implies a variation of the dielectric constant with temperature; however, the implications of this dependence in the behavior of a capacitor are not analyzed. In this article we try to partially amend that omission. The system under study will be a capacitor filled with a polar dielectric that follows a close cycle between a cold and a hot thermal reservoirs; it is found that there is a net energy gain because some heat is transformed to electrostatic potential energy. In addition, this problem can be seen from a thermodynamics point of view: the polar dielectric acts as the "working fluid" in a thermal machine that operates between both heat reservoirs. The analysis of the problem is kept at a level suitable for students of an undergraduate general physics or thermodynamics course.

The structure of this article is the following. Some basic concepts related to dielectrics are briefly reviewed and some equations connected with capacitors are recalled in order to make the paper self-contained. Then, the energy balance of a capacitor filled with a polar dielectric is analyzed while the device performs a closed cycle between a cold and a hot thermal reservoirs. Finally, the system is studied as a thermal engine.

# 2. Dielectrics: a very brief summary

Let us remember some basic concepts about dielectric materials.

1. The dielectric permittivity of a material is called $\varepsilon$ and its related to the vacuum permittivity $\varepsilon_0$ by $\varepsilon = K \varepsilon_0$. $K$ is known as the dielectric constant and it is a purely phenomenological magnitude that has been measured for many materials [7].

2. The constitutive equation of a dielectric is $P = \chi(E)\, \varepsilon_0 E$, where $E$ is the applied electric field, $P$ the material polarizability and $\chi(E)$ the electrical susceptibility that contains microscopic information about the material; we assume it does not depend on $E$, i.e. $\chi(E)=\chi$.
3. The relation between the dielectric constant, the susceptibility and the permittivity is: $K = 1 + \chi = \varepsilon / \varepsilon_0$.
4. The molecules of a dielectric material can be polar or non-polar. A molecule is polar if it has an intrinsic dipole moment $p_0$ that results from an asymmetric charge distribution. Otherwise, the molecule is non-polar.
5. If an external electric field is applied the molecules are polarized, i.e. the field slightly shifts in opposite directions the positive and negatives charges and a dipole moment is induced. If the molecules are non polar, the relation between the external field and the molecular polarization $p_m$ is: $p_m = \alpha\, \varepsilon_0 E$, where $\alpha$ is a proportionality constant. After some [simple] considerations [4, 5], the Clausius-Mossotti equation that relates a microscopic parameter ($\alpha$) with a phenomenological constant ($K$) is obtained:

$$\alpha = \frac{3}{4}\pi N_A \frac{K\varepsilon_0 - 1}{K\varepsilon_0 - 2}, \qquad (1)$$

where $N_{Av}$ is the Avogadro number [, i.e. the number of molecules in one mol].
6. If an external electric field is applied and the material molecules are polar, i.e. they have an intrinsic dipole moment $p_0$, the field tries to align them while the thermal fluctuations disorder the alignment. This situation is well described by the Langevin – Debye equation that gives the value of the constant $\alpha$. There are two terms in the formula. One, called $\alpha_0$, is associated with the modification of the charge distribution induced by the field. The other term takes into account the contribution of the permanent dipole moment and the thermal fluctuations. If $R$ is the gas constant and $T$ the absolute temperature, $\alpha$ is given by:

$$\alpha = \alpha_0 + \frac{p_0^2}{3RT}. \qquad (2)$$

Equation (2) is relevant for our purpose. Comparing the material polarizability $\chi$ and the microscopic proportionality constant $\alpha$, one gets $\chi = N_{Av}\,\alpha$. Taken into account eq. (2) and the relation between $K$ and $\chi$, the temperature dependence of the dielectric constant is:

$$K(T) = A + \frac{B}{T}. \tag{3}$$

The constants $A$ and $B$ are determined experimentally and depend on the dielectric material under study.

## 2. Capacitor energy and its changes

Our next step is to analyze the effect of (3) in the energy $U$ stored in a capacitor. It is well-known that:

$$U = \frac{1}{2}\frac{q^2}{C}. \tag{4}$$

As usual $q$ is the charge and $C$ the capacitance. For a parallel-plates capacitor filled with a dielectric characterized by a permittivity $\varepsilon$, plate area $s$ and distance $d$ between them, the capacitance is:

$$\begin{aligned} C &= \frac{\varepsilon\, s}{d}, \\ &= K\varepsilon_0 \frac{s}{d}, \\ &= M + \frac{N}{T}. \end{aligned} \tag{5}$$

The constants $M$ and $N$ that appears above are easily obtained from the (3):

$$\begin{aligned} M &= A\,\varepsilon_0 s/d, \\ N &= AB\varepsilon_0 s/d. \end{aligned} \tag{6}$$

If the capacitor is not a parallel-plates, it is assumed that the temperature dependence of the capacitance is well described by the last line of (5). In this case, $M$ and $N$ have to be determined experimentally for a given capacitor. The task is simple: measure the capacitance at different temperatures and plot it against *1/T*.

## 3. Capacitor energy and its changes with temperature

The next step is to study the influence of (3) in the energy changes of a capacitor that follows the closed cycle shown in Figure 1a. The capacitor is initially without charge at a temperature $T_c$. It is connected to a battery and receives a charge $q_{max}$. This value is bounded by the dielectric strength of the material [7]. The capacitor is disconnected from the battery and put into contact with a heat reservoir at temperature $T_h$. Since the capacitor is isolated, its charge remains constant. Once the capacitor is at $T_h$, it is discharged through a circuit and useful electric work can be obtained. . Finally the device, with $q=0$, is cooled down to the initial temperature $T_c$. Once hopes that the energy given by the capacitor when it is discharged at $T_h$, would be greater than the energy spent in charging the device at $T_c$. In the next paragraph this calculation is carried out.

Using the last line of eq. (5) and noticing that the temperature correction to the capacitance is small ($N/T << M$), the temperature dependence of voltage and energy can be evaluated:

$$V = \frac{q}{C(T)},$$
$$\cong q\left(M - \frac{N}{T}\right). \quad (7)$$

And:

$$U = \frac{q^2}{2C(T)},$$
$$\cong \frac{q^2}{2}\left(M - \frac{N}{T}\right). \quad (8)$$

In Figures 1b and 1c, $V$ and $U$ are shown as the capacitor follows the cycle described above.

It should be remarked that the energy taken from the capacitor at $T = T_h$ is greater than the given energy at $T = T_c$. The gain $\Delta U$ is:

$$\Delta U = U(T_h) - U(T_c)$$
$$= \frac{q_{max}^2 N}{2}\left(\frac{1}{T_c} - \frac{1}{T_h}\right). \quad (9)$$

Where does this energy gain come from? It comes from the heat absorbed when the capacitor is heated from $T_c$ to $T_h$. The dielectric converts thermal energy into electrostatic one. In thermodynamics says that the capacitor acts as a heat engine that works between two thermal reservoirs at $T_h$ and $T_c$ and the dielectric as the working fluid.

## 4. The capacitors as a heat engine

The capacitor as a heat engine is shown in Figure 2. Heat is absorbed $Q_{ads}$ from the hot reservoir, some useful work $W$ is obtained ($W = \Delta U$) and some heat $Q_{rel}$ is released to the cold reservoir.

The heat absorbed by the "working fluid", i.e. the dielectric, is:

$$Q_{ads} = C_{th}(T_h - T_c). \tag{10}$$

$C_{th}$ is the heat capacity of the dielectric. If the capacitor is parallel-plates, $C_V$ is the specific heat of the dielectric that can be taken as a constant in the considered temperature interval and $\rho$ is the material density, then the heat capacity $C_{th}$ of the dielectric that fills the gap in a parallel-plates capacitor is:

$$C_{th} = s\,d\,\rho\,C_V. \tag{11}$$

The heat transferred to the cold reservoir in the cooling branch of the process is:

$$Q_{rel} = -\left(C_{th}(T_h - T_c) - \Delta U\right). \tag{12}$$

The efficiency $\eta$ of the heat engine results from (9) and (10):

$$\begin{aligned}\eta &= \frac{\Delta U}{Q_{ads}} \\ &= \frac{1}{2}\frac{q_{max}^2 N}{C_{th}}\frac{1}{T_h T_c}.\end{aligned} \tag{13}$$

It is clear that the energy gain (9) and the engine efficiency (13) can be evaluated from two parameters – $N$ and $C_{th}$ – than can be measured for a given capacitor. If the capacitor can be considered as parallel-plates, then $N$ and $C_{th}$ can be calculated from known properties of the dielectric –eqs. (5) y (11)–.

## 5. A suggestion for further work and summary

There might be a technological application for the analysis carried out in this paper. Solar energy can be used to heat up the dielectric, and electrical energy can be obtained without using solar cells. However, it is hard to say how efficient this conversion would be.

Experimental work with real capacitors is needed to determine $N$ and $C_{th}$ and the possible values of $q_{max}$ that are bounded by the dielectric strength.

In summary, this article shows how the temperature dependence of the dielectric constant entails a surprising result with the capacitor energy. The article is kept at a level suitable for students who take a general physics course with a brief introduction to the microscopic theory of dielectrics. Additionally, this physics exercise can be used in a thermodynamics course to show that the working fluid in a heat engine can be a dielectric. Finally, the possible use of a dielectric to convert thermal to electrical energy might deserve attention.

**Acknowledgment:** The author thanks the National Scientific and Technological Council (CONICET) of Argentina for financial support.

# Figure captions

**Figure 1:**

(a) Cycle followed by a capacitor filled with a dielectric with a temperature dependent permittivity. Initially the capacitor has no charge and is in contact with a cold reservoir at temperature $T_c$. Then the device is connected to a battery and it charges until $q_{max}$. Afterwards, the capacitor is disconnected from the battery and put into contact with a hot reservoir at temperature $T_h$. At this temperature, the capacitor discharges through a circuit and some useful work may be obtained. Once the device has no charge, it is cooled down to the initial temperature $T_c$.

(b) The evolution of the voltage V between the capacitor plates is show while the cycle described above is followed.

(c) The capacitor energy $U$ shown in terms of the temperature while the capacitor performs the cycle. Notice that the energy at $T_h$ is greater than the energy at $T_c$. The dielectric converts thermal to electrostatic energy that is stored in the capacitor.

**Figure 2:**

The capacitor, seen from a thermodynamics point on view, behaves as a heat engine that absorbs heat $Q_{ads}$ from a hot reservoir at $T_h$, converts some heat into usable work $W = \Delta U$ –see eq. (8)– and releases some heat $Q_{rel}$ to a cold reservoir at $T_c$. From energy conservation, $Q_{rel} = Q_{ads} - W$.

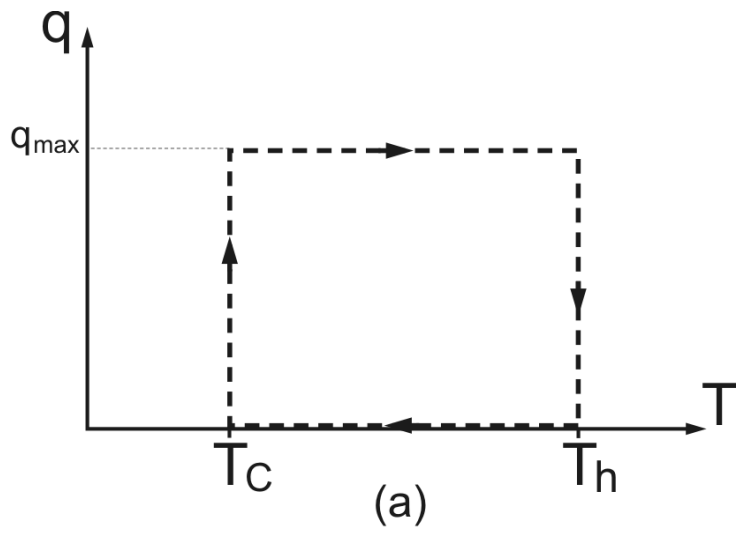

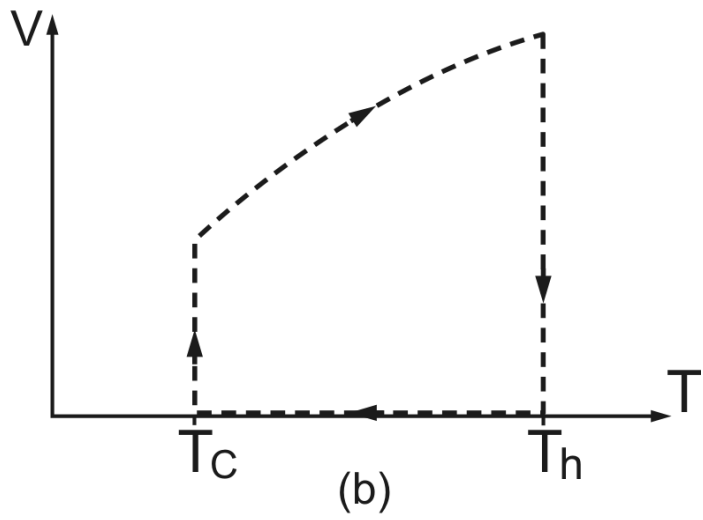

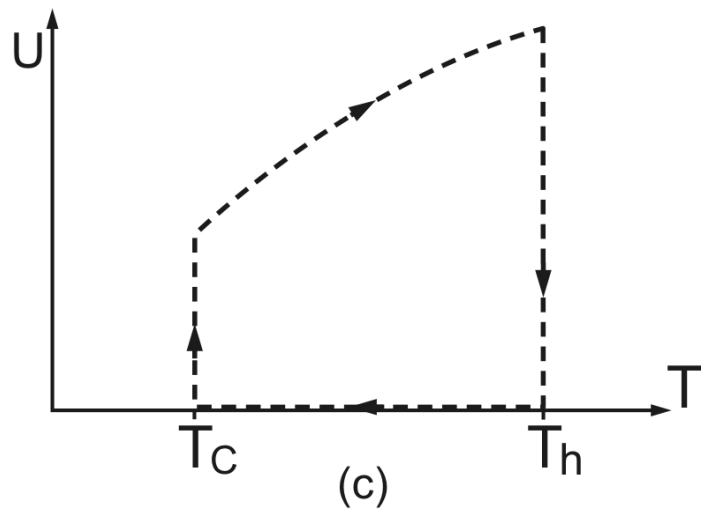

Figure 1

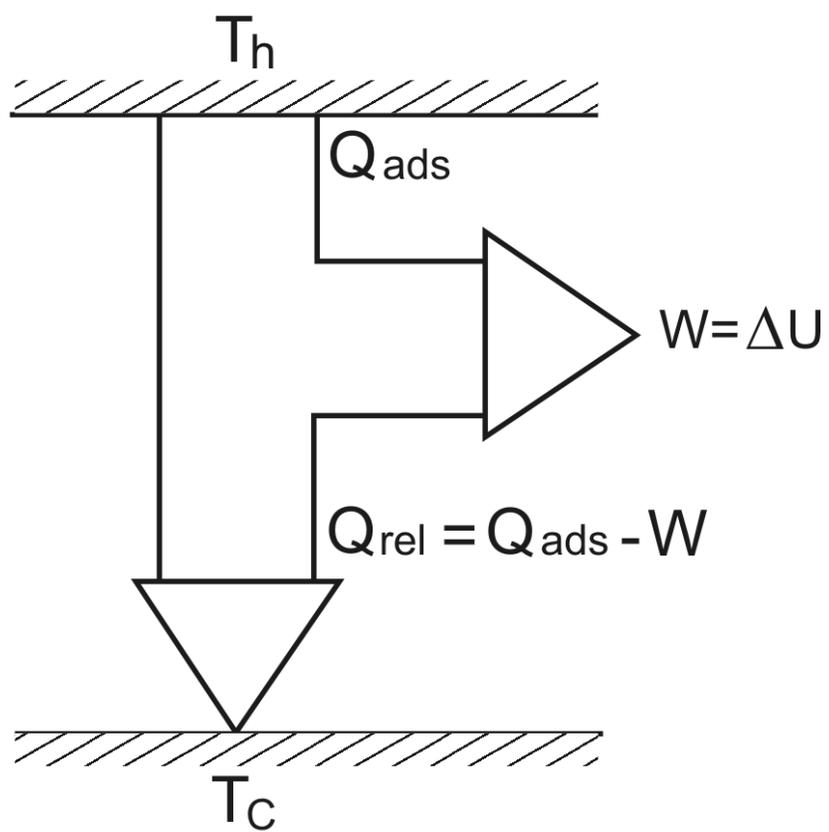

Figure 2